\documentstyle[twoside,epsf]{article}

\input{ibvs2.sty}

\begin{document}

\IBVShead{xxxx}{xx xxx 2002}

\IBVStitletl{Astrometric authentication of RX J2309.8+2135}{as a nearby dwarf nova candidate}

\IBVSauth{Kato,~Taichi$^1$; Yamaoka,~Hitoshi$^2$}
\vskip 5mm

\IBVSinst{Dept. of Astronomy, Kyoto University, Kyoto 606-8502, Japan,
          e-mail: tkato@kusastro.kyoto-u.ac.jp}

\IBVSinst{Faculty of Science, Kyushu University, Fukuoka 810-8560, Japan,
          e-mail: yamaoka@rc.kyushu-u.ac.jp}

\IBVSobj{RX J2309.8+2135}
\IBVStyp{UG:}
\IBVSkey{X-ray source, cataclysmic variable, proper motion, classification}

\begintext

   RX J2309.8+2135 = 1RXS J230949.6+213523 is an X-ray source discovered
through the ROSAT all-sky survey (Voges et al. 1999).  Wei et al. (1999)
reported its optical identification and provided a classification as
a cataclysmic variable (CV) without further specification.
Schwope et al. (2000) tentatively classified this object as a symbiotic
binary based on the presence of a strong M-type stellar component.
Most recently, however, Schwope et al. (2002) reported a refined
classification as a CV based on the complete absence of high ionization
lines and the line ratios (He I and Balmer lines) unlike those of
symbiotic stars.  Schwope et al. (2002) further proposed the spectral
type of M3.  Assuming that the secondary star is a Roche-lobe filling
M3 dwarf star, Schwope et al. (2002) suggested an extremely small
distance of $\sim$30 pc.
This implication is surprising since this would break the nearest
record of CVs (the best established example being WZ Sge: 45 pc,
J. Thorstensen, cited in Steeghs et al. (2001)).  We therefore reexamined
this possibility.

   We have examined the available astrometric catalogs (Table 1),
and detected a large proper motion of 2$''$.8 in 40.1 yr.  This value
has been confirmed by a direct comparison of DSS 1 and 2 plate scans.
The detected proper motion corresponds to 0$''$.069$\pm$0$''$.012
yr$^{-1}$.  This value is comparable to nearby dwarf novae with
large proper motions (WZ Sge: 0$''$.078$\pm$0$''$.007 yr$^{-1}$,
Kraft and Luyten 1965; GW Lib: 0$''$.066$\pm$0$''$.012 yr$^{-1}$,
Thorstensen et al. 2002; V893 Sco: 0$''$.067$\pm$0$''$.015 yr$^{-1}$,
Thorstensen 1999; 1RXS J232953.9+062814: 0$''$.056$\pm$0$''$.005 yr$^{-1}$,
Uemura et al. 2001, Kimeswenger et al. 2002).  This large proper
motion makes the object likely a nearby object.  Assuming a upper limit
transverse velocity of 100 km s$^{-1}$, the upper limit of the distance
becomes $\sim$300 pc, which is consistent with independent distance
determinations of CVs with similar proper motions (Thorstensen et al. 2002).
The upper limit of $M_{\rm V}$ of the secondary thus becomes +8.5,
which safely excludes the possibility of a symbiotic binary with a
giant secondary.  This indication is consistent with the suggested
luminosity classification based on the CaH absorption at 6382\,\AA
(Schwope et al. 2002).
The present astrometry thus authenticates RX J2309.8+2135 as a nearby
dwarf nova candidate.  Even at a reasonable distance of $\sim$100 pc,
the outburst maxima of RX J2309.8+2135 would reach $V \sim$9.  At the
suggested distance of $\sim$30 pc, the maxima would reach even $V \sim$7
(Warner 1986).
These values indicate that RX J2309.8+2135 is a candidate for the brightest
dwarf novae in the entire sky.

\begin{table}
\begin{center}
Table 1. Astrometry of RX J2309.8+2135. \\
\vspace{10pt}
\begin{tabular}{cccc}
\hline\hline
Source    & R. A. & Decl.               & Epoch \\
          & \multicolumn{2}{c}{(J2000.0)} & \\
\hline
USNO A2.0 & 23 09 49.27 & +21 35 20.0 & 1951.613 \\
GSC 2.2.1 & 23 09 49.17 & +21 35 17.7 & 1991.754 \\
\hline
\end{tabular}
\end{center}
\end{table}

   However, the observed properties of RX J2309.8+2135 is rather unusual
for a typical dwarf nova.  The spectral type of M3 usually indicates an
orbital period longer than 2--3 hr.  Although CVs with such periods
usually have relatively high mass-transfer rates (Warner 1995), the weak
contribution of a disk continuum in the published spectra of
RX J2309.8+2135 suggests the contrary.
The lack of outburst detection between 2002 January and October (VSNET
observations) seems to support a low mass-transfer rate.  We know another
secondary-dominated system with a low mass-transfer rate, CW 1045+525
(Tappert et al. 2001).  These objects may represent a hitherto unidentified
class of long-period CVs with the lowest mass-transfer rates, or these
system may be undergoing excursions to long-lasting low states as in
VY Scl-type stars (Warner 1995).  The rather narrow appearance of emission
lines in the published spectra resembles those of low states in VY Scl-type
stars (Robinson et al. 1981), although a nearly pole-on view of a dwarf
nova is also consistent with the observation.  Since the orbital period of
RX J2309.8+2135 has not yet been established, there also remains a
possibility of an object with an anomalously hot, bright, evolved secondary
(Uemura et al. 2002).  The object is a very good candidate for future
detailed photometric and radial velocity studies.

\vskip 3mm

We are grateful to Pavol A. Dubovsky and Timo Kinnunen who reported
observations to VSNET.
This work is partly supported by a grant-in aid [13640239 (TK),
14740131 (HY)] from the Japanese Ministry of Education, Culture, Sports,
Science and Technology.

\references

Kimeswenger, S., Schmeja, S., Kitzbichler, M. G., Lechner, M. F. M.,
   M\"{u}hlbacher, M. S.; M\"{u}hlbauer, A. D., 2002, IBVS No. 5233

Kraft, R. P., Luyten, W. J., 1965, ApJ, {\bf 142}, 1041

Robinson, E. L., Barker, E. S., Cochran, A. L., Cochran, W. D.,
   Nather, R. E., 1981, ApJ, {\bf 251}, 611

Schwope, A. D., Hasinger, G., Lehmann, I., Schwarz, R., Brunner, H.,
   Neizvestny, S., Astron. Nachr. {\bf 321}, 1

Schwope, A. D., Brunner, H., Buckley, D., Greiner, J., van der Heyden, K.
   et al., 2002, A\&A, in press (astro-ph/0210059)

Steeghs, D., Marsh, T., Knigge, C., Maxted, P. F. L., Kuulkers, E.,
   Skidmore, W., 2001, ApJ, {\bf 562}, L145

Tappert, C., Thorstensen, J. R., Fenton, W. H., Bennert, N.,
   Schmidtobreick, L., Bianchini, A., 2001, A\&A, {\bf 380}, 533

Thorstensen, J. R., 1999, IBVS No. 4749

Thorstensen, J. R., Patterson, J. O., Kemp, J., Vennes, S., 2002,
   PASP, in press (astro-ph/0206426)

Uemura, M., Ishioka, R., Kato, T., Schmeer, P., Yamaoka, H., Starkey, D.
   et al., 2001, IAUC No. 7747

Uemura, M., Kato, T., Ishioka, R., Yamaoka, H., Schmeer, P., Starkey, D. R.
   et al., 2002, PASJ, {\bf 54}, L15

Voges, W., Aschenbach, B., Boller, T., Braeuninger, H., Briel, U.,
   Burkert, W. et al., 1999, A\&A, {\bf 349}, 389

Warner, B., 1986, MNRAS, {\bf 222}, 11

Warner, B., 1995, Cataclysmic Variable Stars (Cambridge: Cambridge
  University Press)

Wei, J. Y., Xu, D. W., Dong, X. Y., Hu, J. Y., 1999, A\&AS, {\bf 139}, 575

\endreferences

\end{document}